\documentclass[12pt]{article}

\usepackage{epsf}
\usepackage{latexsym,amssymb,euscript}
\usepackage[dvips]{graphicx}
\usepackage{amsmath}

\begin{document}

\begin{center}
{\Large \textbf{Phase structure of $3D$ $Z(N)$ lattice gauge theories 
at finite temperature}}
\end{center}

\vskip 0.3cm
\centerline{O.~Borisenko$^{1\dagger}$, V.~Chelnokov$^{1*}$, 
G.~Cortese$^{2\dagger\dagger}$, M.~Gravina$^{3\ddagger}$, A.~Papa$^{4\P}$, 
I.~Surzhikov$^{1**}$}

\vskip 0.6cm

\centerline{${}^1$ \sl Bogolyubov Institute for Theoretical Physics,}
\centerline{\sl National Academy of Sciences of Ukraine,}
\centerline{\sl 03680 Kiev, Ukraine}

\vskip 0.2cm

\centerline{${}^2$ \sl Instituto de F\'{\i}sica Te\'orica UAM/CSIC,}
\centerline{\sl Cantoblanco, E-28049 Madrid, Spain}
\centerline{\sl and Departamento de F\'{\i}sica Te\'orica,}
\centerline{\sl Universidad de Zaragoza, E-50009 Zaragoza, Spain}

\vskip 0.2cm

\centerline{${}^4$ \sl Department of Physics, University of Cyprus,
P.O. Box 20357, Nicosia, Cyprus}

\vskip 0.2cm

\centerline{${}^3$ \sl Dipartimento di Fisica, Universit\`a della 
Calabria,}
\centerline{\sl and Istituto Nazionale di Fisica Nucleare, 
Gruppo collegato di Cosenza}
\centerline{\sl I-87036 Arcavacata di Rende, Cosenza, Italy}

\vskip 0.6cm

\begin{abstract}
We perform a numerical study of the phase transitions in three-dimensional 
$Z(N)$ lattice gauge theories at finite temperature for $N>4$. Using the
dual formulation of the models and a cluster algorithm we locate 
the position of the critical points and study the critical behavior across 
both phase transitions in details. In particular, we determine various 
critical indices, compute the average action and the specific heat. Our 
results are consistent with the two transitions being of infinite order. 
Furthermore, they belong to the universality class of two-dimensional 
$Z(N)$ vector spin models.  
\end{abstract}

\vfill
\hrule
\vspace{0.3cm}
{\it e-mail addresses}:

$^\dagger$oleg@bitp.kiev.ua, \ \ $^*$vchelnokov@i.ua,
\ \ $^{\dagger\dagger}$cortese@unizar.es,
\ \ $^{\ddagger}$gravina@ucy.ac.cy,

$^{\P}$papa@cs.infn.it, \ \ $^{**}$i\_van\_go@inbox.ru

\newpage

\section{Introduction}

The deconfinement phase transition in finite-temperature lattice gauge 
theories (LGTs) has been one of the main subjects of investigation for the 
last three decades. 
By now it is well studied and understood for a number of pure gauge models 
in dimensions $D=3,4$. In particular, the phase structure of a 
finite-temperature three-dimensional ($3D$) pure $SU(N)$ LGT with the standard 
Wilson action is thoroughly investigated both for $N=2,3$ and for the 
large-$N$ limit (see, {\it e.g.},~\cite{3D_sun} and references therein). The 
transition is second order for $N=2,3$ and first order for $N>4$. In the case 
of the $SU(4)$ gauge group, most works agree that the transition is weakly 
first order. The deconfining transition in $SU(N=2,3)$ LGTs belongs to the 
universality class of $2D$ $Z(N=2,3)$ Potts models.  
All these phase transitions are characterized by the spontaneous symmetry 
breaking of a $Z(N)$ global symmetry of the lattice action in the 
high-temperature deconfining phase. 

Another interesting set of models is represented by abelian $Z(N)$ LGTs. 
$Z(N)$ is the center subgroup of $SU(N)$, hence $Z(N)$ LGT can provide 
useful insights into the universal properties of $SU(N)$ models. 
Moreover, $Z(N)$ LGTs are interesting on their own right and might 
possess an even richer phase structure as will be revealed below.   
The most general action for the $Z(N)$ LGT can be written as 
\begin{equation} 
S_{\rm gauge} \ = \ \sum_x \sum_{n<m} \ \sum_{k=1}^N \beta_k
\cos \left( \frac{2 \pi k}{N} \left(s_n(x) + s_m(x+e_n) 
-s_n(x+e_m) - s_m(x) \right) \right) \ .
\label{action_gauge}
\end{equation}
Gauge fields are defined on links of the lattice and take on values 
$s_n(x)=0,1,\cdots,N-1$. $Z(N)$ gauge models, similarly to their spin cousins, 
can generally be divided into two classes - the standard Potts models and the 
vector models. 
The standard gauge Potts model corresponds to the choice when all 
$\beta_k$ are equal. Then, the sum over $k$ in~(\ref{action_gauge}) reduces 
to a delta-function on the $Z(N)$ group. The conventional vector model 
corresponds to $\beta_k=0$ for all $k>1$. For $N=2,3$ the Potts and vector 
models are equivalent. 

The study of the phase structure of $Z(N)$ LGTs at zero temperature has a long 
history. While the phase structure of the general model defined 
by~(\ref{action_gauge}) remains unknown, it is well established that the 
Potts models and vector models with only $\beta_1$ non-vanishing have one 
phase transition from a confining phase to a phase with vanishing string 
tension~\cite{horn,ukawa,savit}. 
Via duality, $Z(N)$ gauge models can be exactly related to $3D$ $Z(N)$ spin 
models. In particular, a Potts gauge theory is mapped to a Potts spin model, 
and such a relation allows to establish the order of the phase 
transition. Hence, Potts LGTs with $N=2$ have second order phase transition, 
while for $N\geq 3$ one finds a first order phase transition. 
Vector models have been studied numerically in~\cite{bhanot} up to $N=20$.
It was confirmed that the zero-temperature models possess a single phase 
transition which disappears in the limit $N\to\infty$. Thus, the $U(1)$ LGT 
has a single confined phase in agreement with theoretical 
results~\cite{3d_u1}. A scaling formula for the critical coupling with $N$ 
had also been proposed in~\cite{bhanot}. 
We are not aware, however, of any detailed study of the critical behavior 
of the vector models with $N>4$ in the vicinity of this single phase 
transition. 

The deconfinement phase transition at finite temperature is well understood 
and studied for $N=2,3$. An especially detailed study was performed on the 
gauge Ising model, $N=2$, in~\cite{3D_z2_potts}.  
These models belong to the universality class of $2D$ $Z(N)$ spin models and 
exhibit a second order phase transition in agreement with the Svetitsky-Yaffe 
conjecture~\cite{svetitsky}. 
One should expect on general grounds that the gauge Potts models possess a 
first order phase transition for all $N> 4$, similarly to $2D$ Potts models. 
The $Z(4)$ vector model has been simulated, {\it e.g.}, in~\cite{3D_z4_at}. It 
also belongs to the universality class of the $2D$ $Z(4)$ spin model and 
exhibits a second order transition. 

Much less is known about the finite-temperature deconfinement transition 
for the vector $Z(N)$ LGTs when $N>4$. It is the ultimate goal of the present 
work to deepen our understanding of the phase structure of these models. 
The Svetitsky-Yaffe conjecture is known to connect critical properties 
of $3D$ $Z(N)$ LGTs at finite temperature with the corresponding 
properties of $2D$ spin models, if they share the same global symmetry 
of the action. It is widely expected, 
and in many cases proved by either analytical or numerical methods, that some 
$2D$ $Z(N>4)$ spin models (like the vector Potts model) have two phase 
transitions of infinite order, known as the Berezinskii-Kosterlitz-Thouless 
(BKT) phase transitions. According to the conjecture, the phase 
transitions in some $3D$ $Z(N>4)$ gauge models at finite temperature could 
exhibit two phase transitions as well. 
Moreover, if the correlation length diverges when approaching the critical 
point, these transitions should be of the BKT type and belong to the 
universality class of the corresponding $2D$ $Z(N)$ spin models.

The BKT phase transition has been best studied in the $2D$ $XY$ 
model~\cite{berezin,kosterlitz1,kosterlitz2}. 
Certain analytical~\cite{svetitsky,parga,lat_07} 
as well as numerical results~\cite{3du1ft,3du1full} unambiguously indicate 
that the deconfining phase transition in the $3D$ $U(1)$ LGT is also of 
infinite order and might belong to the universality class of $2D$ $XY$ 
model~\footnote{It should be noted, however, that the numerical results 
of~\cite{3du1full} point to a critical index~$\eta$ larger than its $XY$ 
value by almost a factor of 2 for $N_t=8$. Therefore, the question of the 
universality remains open for this model.}. 

In recent papers~\cite{3d_zn_strcoupl,lat_12} we have initiated exploring the 
phase structure of the vector $Z(N)$ LGT for $N>4$. More precisely, we have 
considered an anisotropic lattice in the limit where the spatial coupling 
vanishes. In this limit the spatial gauge fields can be exactly integrated out 
and one gets a $2D$ generalized $Z(N)$ model. The Polyakov loops play the 
role of $Z(N)$ spins in this model. For the Villain version of the model 
obtained we have been able to present renormalization group arguments 
indicating the existence of two BKT-like phase transitions. This scenario was 
confirmed with the help of large-scale Monte Carlo simulations of the 
effective model. 
We have also computed some critical indices which appear to agree with 
the corresponding indices of $2D$ $Z(N)$ spin models, thus giving 
further support to the Svetitsky-Yaffe conjecture. 

In this paper we extend our analysis to the full isotropic $3D$ $Z(N)$ LGT  
at finite temperature. It is well known that the full phase structure 
of a finite-temperature LGT is correctly reproduced in the limit where 
spatial plaquettes are neglected. They have probably small influence 
on the dynamics of the Polyakov loop interaction.  
We therefore expect that the scenario advocated by us in~\cite{3d_zn_strcoupl} 
remains qualitatively correct for the full theory. 
In particular, full gauge models with $N>4$ should possess two phase 
transitions of the BKT type and we expect the values of critical indices to 
coincide with the indices of the $2D$ vector spin models. 

The $2D$ $Z(N)$ spin model in the Villain formulation has been studied 
analytically in Refs.~\cite{savit,elitzur,kogut,nienhuis,kadanoff,Cardy}. It 
was shown that the model has at least two phase transitions when $N\geq 5$. 
The intermediate phase is a massless phase with power-like decay of the 
correlation function. 
It turns out that $\eta(\beta^{(1)}_{\rm c})=1/4$ at the transition point 
from the strong coupling high-temperature phase to the massless phase, 
{\it i.e.} the behavior is similar to that of the $XY$ model. At the 
transition point $\beta^{(2)}_{\rm c}$ from the massless phase to the ordered 
low-temperature phase one has $\eta(\beta^{(2)}_{\rm c})=4/N^2$.
A rigorous proof that the BKT phase transition does take place, and so that the
massless phase exists, has been constructed in Ref.~\cite{rigbkt} for both
Villain and standard formulations of the vector model. 
Universality properties of vector models were studied via Monte Carlo 
simulations in Ref.~\cite{cluster2d} for $N=6,8,12$ and in 
Refs.~\cite{lat_10,2dzn,lat_11,2dzn7_17} for $N=5,7,17$. 
Results for the critical indices $\eta$ and $\nu$ agree with 
analytical predictions obtained for the Villain formulation of the model.

A similar phase structure is expected to hold for the finite-temperature 
$3D$ $Z(N>4)$ LGT. It can be described in terms of the Polyakov loop 
correlation functions as follows. The low-temperature phase is a confining 
phase. Here, the correlation decays with an area law, thus implying a 
non-vanishing string tension and a linear potential between static 
$Z(N)$ charges. With the temperature increasing, the system undergoes a phase 
transition to a massless phase. This phase is characterized by the enhancement 
of the $Z(N)$ global symmetry to a $U(1)$ symmetry and is very close in nature 
to the high-temperature phase of $U(1)$ gauge model. In particular, the 
dominating contribution to the correlation of the Polyakov loops comes from 
massless excitations (dual spin-waves). This is nevertheless a confinement 
phase, as the correlation decays with a power law and though the string 
tension vanishes, the potential between $Z(N)$ charges is logarithmic. 
Increasing the temperature further on leads to a spontaneous breaking of the
$Z(N)$ global symmetry at some critical point. 
One enters a deconfining phase above this critical point.  

Here we intend to: 

\begin{itemize} 

\item 
check the scenario of two BKT phase transitions described above; 

\item 
compute some critical indices at both transitions and verify the universality 
class of the model. 

\end{itemize}

The fact that the BKT transition has infinite order makes it hard to study 
its properties using analytical methods. In most of the cases studied so far
one uses a renormalization group (RG) technique as in Ref.~\cite{elitzur}. 
Unfortunately, there are no direct ways to generalize the transformations 
of Ref.~\cite{elitzur}, leading to RG equations, to $3D$ $Z(N)$ LGTs, except 
for the limiting case $N\to\infty$. To study the phase structure of these 
models we need numerical simulations. Here, however, another problem appears
related to the very slow, logarithmic convergence to the thermodynamic limit 
in the vicinity of the BKT transition. It is thus necessary to use both 
large-scale simulations and combine them with the finite-size scaling methods. 
Our principal strategy consists in passing to a dual formulation of the 
$3D$ $Z(N)$ vector LGT, which is known to be a generalized $3D$ $Z(N)$ vector 
model. This allows us to use a cluster algorithm in our simulations. 
The standard procedure in studying the phase structure is to use 
Binder cumulants to locate the position of critical points. Then, critical 
indices can be determined from various susceptibilities. In the case of a 
deconfinement phase transition both Binder cumulants and susceptibilities are 
usually constructed from Polyakov loops. However, it is a non-trivial problem 
to write down the expression for a single Polyakov loop in a dual formulation 
(though the dual form can be easily found for invariant quantities, like 
the correlation of Polyakov loops). 
We have therefore decided to study the critical behavior making use of 
Binder cumulants and susceptibilities constructed from the dual $Z(N)$ 
spins~\footnote{The use of the Polyakov loop correlations to directly extract 
critical indices requires, in case of an essential singularity occurring at 
the BKT transition, prohibitively huge lattices, which are not accessible with 
the present numerical facilities.}. This procedure exhibits an 
interesting phenomenon, namely the critical behavior of dual spins is 
reversed with respect to the critical behavior of Polyakov loops: 
the spontaneously-broken ordered phase is mapped to the symmetric phase and 
{\it vice versa}. Moreover, the critical indices $\eta$ are also interchanged 
as will be explained later on. The index $\nu$ which governs the exponential 
divergence of the correlation length is expected to be the same at both 
transitions and takes on the value $\nu=1/2$. 

The lowest number of $N$ where two BKT phase transitions are expected is 
$N=5$. In the present paper we study the phase transitions in models with 
$N=5,13$ in great detail. We have studied these values of $N$ in the strong 
coupling regime~\cite{3d_zn_strcoupl,lat_12}, therefore it is natural to
continue working with these models. Our computations are performed on lattices 
with temporal extent $N_t=2,4$ and with spatial size in the range 
$L\in [32-1024]$. 

This paper is organized as follows. In Section~2 we formulate our model and 
establish the exact relation with a generalized $3D$ $Z(N)$ spin model. 
Here, we also explain why critical indices, determined from the dual spin 
correlation functions, become $\eta(\beta^{(1)}_{\rm c})=4/N^2$ and 
$\eta(\beta^{(2)}_{\rm c})=1/4$, {\it i.e.} the values are interchanged 
with the respect to what expected from the correlation of  
original degrees of freedom. In Section~3 we present
the setup of Monte Carlo simulations, define the observables used in this work
and present the numerical results of simulations. In particular, we locate the 
position of critical points and compute various critical indices at these 
points. As a further cross-check of the nature of the phase transitions, we 
also present results of a few simulations for the $Z(5)$ gauge model and 
compute the average action and the specific heat in the vicinity of critical 
points. Our conclusions and perspectives are given in Section~5.   

\section{Relation of the $3D$ $Z(N)$ LGT to a generalized $3D$ $Z(N)$ spin model} 

\subsection{Partition and correlation functions} 

We work on a $3D$ lattice $\Lambda = L^2\times N_t$ with spatial extension $L$ 
and temporal extension $N_t$; $\vec{x}=(x_0,x_1,x_2)$, where $x_0\in [0,N_t-1]$
and $x_1,x_2\in [0,L-1]$ denote the sites of the lattice and $e_n$, $n=0,1,2$, 
denotes a unit vector in the $n$-th direction.
Periodic boundary conditions (BC) on gauge fields are imposed in all 
directions. The notations $p_t$ ($p_s$) stand for the temporal (spatial) 
plaquettes, $l_t$ ($l_s$) for the temporal (spatial) links.

We introduce conventional plaquette angles $s(p)$ as
\begin{equation}
s(p) \ = \ s_n(x) + s_m(x+e_n) - s_n(x+e_m) - s_m(x) \ .
\label{plaqangle}
\end{equation}
The $3D$ $Z(N)$ gauge theory on an anisotropic lattice can generally be 
defined as 
\begin{equation}
Z(\Lambda ;\beta_t,\beta_s;N) \ = \  \prod_{l\in \Lambda}
\left ( \frac{1}{N} \sum_{s(l)=0}^{N-1} \right ) \ \prod_{p_s} Q(s(p_s)) \
\prod_{p_t} Q(s(p_t)) \ .
\label{PTdef}
\end{equation}
The most general $Z(N)$-invariant Boltzmann weight with $N-1$ different 
couplings is
\begin{equation}
Q(s) \ = \
\exp \left [ \sum_{k=1}^{N-1} \beta_p(k) \cos\frac{2\pi k}{N}s \right ] \ .
\label{Qpgen}
\end{equation}
The Wilson action corresponds to the choice $\beta_p(1)=\beta_p$, 
$\beta_p(k)=0, k=2,...,N-1$. The $U(1)$ gauge model is defined as the limit 
$N\to\infty$ of the above expressions.

In what follows we work with the conventional Wilson action on an isotropic 
lattice, {\it i.e.} $\beta_s=\beta_t=\beta$. 
To study the phase structure of $3D$ $Z(N)$ LGTs one can map the gauge model 
to a generalized $3D$ spin $Z(N)$ model with the action 
\begin{equation}
\label{modaction}
S \ =\ \sum_{x}\ \sum_{n=1}^3 \sum_{k = 1}^{N-1} \ \beta_k \  
\cos \left( \frac{2 \pi k}{N} \left(s(x) - s(x+e_n) \right) \right) \ .
\end{equation}
The effective coupling constants $\beta_k$ can be computed exactly as follows. 
The first step is to construct a dual form for the partition 
function~(\ref{PTdef}). Details can be found, {\it e.g.}, in~\cite{ukawa}. 
One gets on the dual lattice $\Lambda_d$ the following expression 
for the partition function~\footnote{In writing this expression we have 
neglected a certain global summation which appears due to the fact that the 
product of the plaquette variables around the closed $2d$ surface winding 
through the lattice in periodic directions is unity (so-called global Bianchi 
identity). Such global identities can be safely omitted since they do not 
influence the quantities of our interest in thermodynamic limit. 
Note, this is not the case for quantities like the twist free energy.}
\begin{equation}
Z(\Lambda_d ;\beta ;N) \ = \  \prod_{x\in \Lambda_d}
\left ( \frac{1}{N} \sum_{s(x)=0}^{N-1} \right ) \ \prod_{l\in\Lambda_d} 
\ Q_d(s(x)-s(x+e_n)) \ ,
\label{PTdual}
\end{equation}
where the dual Boltzmann weight $Q_d(s)$ becomes 
\begin{equation}
Q_d(s) \ = \ \sum_{r=-\infty}^{\infty} \ I_{Nr+s}(\beta)  \ = \ 
\sum_{p = 0}^{N - 1} \exp \left[ \beta \cos \left(\frac{2 \pi p}{N} 
\right) \right ] \cos \left(\frac{2 \pi p s}{N} \right) \ .
\label{Qdual}
\end{equation} 
Here, $I_k(x)$ is the modified Bessel function. 
Exponentiating and re-expanding the dual weight in a new Fourier series one 
finds $\beta_k$ as 
\begin{equation}
\beta_k \ =\ \frac{1}{N} \sum_{p = 0}^{N - 1} \ln \left [ \frac{Q_d(p)}
{Q_d(0)} \right ] \  \cos \left(\frac{2 \pi p k}{N} \right) \ .
\label{couplings}
\end{equation}
As example, we give below the expressions for the effective couplings 
$\beta_k$ for $N=5$. One obtains from~(\ref{Qdual}) and~(\ref{couplings}) 
\begin{eqnarray} 
\beta_1 \ &=& \ \beta_4 \ = \ -\frac{1-\sqrt{5}}{10}\ln[1-t_+/2] - 
\frac{1+\sqrt{5}}{10}\ln[1-t_-/2] \ , \  \nonumber   \\ 
\beta_2 \ &=& \ \beta_3 \ = \ -\frac{1+\sqrt{5}}{10}\ln[1-t_+/2] - 
\frac{1-\sqrt{5}}{10}\ln[1-t_-/2] \ ,
\label{betak_N5}
\end{eqnarray}
where
\begin{equation*} 
t_{\pm} \ = \  \frac{5 \pm\sqrt{5}+(5 \mp \sqrt{5})e^{\frac{\sqrt{5}}{2}\beta}}
{2+2e^{\frac{\sqrt{5}}{2}\beta} + e^{\frac{1}{4}(5+\sqrt{5})\beta}} \ \ . 
\end{equation*} 
As is seen from Eq.~(\ref{Qdual}), the dual model is ferromagnetic. However, 
$\beta_2$ is small and negative for $\beta>1.077$. In the whole region 
$| \beta_1 | \gg | \beta_2 |$; moreover, in the critical region 
$| \beta_2 | / | \beta_1 | \sim 10^{-2}$. Similar properties hold for all $N$. 
Thus, one expects that the $3D$ vector spin model with only $\beta_1$ 
non-vanishing gives a reasonable approximation to the gauge model (in our 
simulations we use all $\beta_k$). Next important fact, evident from 
Eq.~(\ref{betak_N5}), is that the weak and the strong coupling regimes are 
interchanged: when $\beta\to\infty$ both effective couplings $\beta_k\to 0$ 
and, therefore, the ordered symmetry-broken phase is mapped to a symmetric 
phase with vanishing magnetization of dual spins. The symmetric phase at small 
$\beta$ becomes an ordered phase where the dual magnetization is non-zero.  

Let $W(x)=\prod_{x_0=0}^{N_t-1}\exp (\frac{2\pi i}{N}j\ s(x))$ be the Polyakov 
loop in the representation $j$. The correlation function of Polyakov loops can 
be expressed in the dual form as 
\begin{equation}
P_j(R;\beta;N) \ = \ Z(\Lambda_d ;\beta ;N)^{-1} \ \prod_{x\in \Lambda_d}
\left ( \frac{1}{N} \sum_{s(x)=0}^{N-1} \right ) \ \prod_{l\in\Lambda_d} 
\ Q_d(s(x)-s(x+e_n)+h(l)) \ .
\label{PLcorr_dual}
\end{equation}
Here we have introduced the sources $h(l)$ as 
\begin{equation}
 h(l) \ = \ 
\begin{cases}
 j, & l\in S_d \ , \ l=(x,n) \\
-j, & l\in S_d \ , \ l=(x-e_n,n) \\
0, & \text{otherwise}
\end{cases}
\label{source}
\end{equation}
where $S_d$ is the dual surface enclosed between two Polyakov loops, 
{\it i.e.} it consists of links perpendicular to plaquettes of the original 
lattice and is closed in the temporal direction. 

The correlation function of the dual spins is defined in standard way as 
\begin{equation}
\Gamma_j(R;\beta;N) \ = \ \left \langle \ 
\exp \left [ \frac{2 \pi i}{N}\ j \ (s(0)-s(R)) \right ] \ \right  \rangle  \ .
\label{Corr_dualspin}
\end{equation}
In the original (gauge) formulation this correlation function corresponds to 
a disorder operator which is the 't Hooft line in the $3D$ theory. 
It consists of a string of plaquettes connecting the cubes dual to the points 
$0$ and $R$ and measures the free energy of a $Z(N)$ monopole--anti-monopole 
pair. 

\subsection{Behavior of the dual correlation function}  

As was explained in the Introduction, our aim is to study the critical 
behavior making use of quantities, like the Binder cumulants, constructed from 
the dual $Z(N)$ spins. On very general grounds one expects that in the 
confined phase, where the potential between electric $Z(N)$ charges grows 
linearly with the distance, the correlation function~(\ref{Corr_dualspin}) 
shows ordered behavior, so that the magnetic charges are free. In the phase 
where the electric charges are deconfined, the correlation 
function~(\ref{Corr_dualspin}) exhibits exponential decay and the free energy 
of a monopole--anti-monopole pair grows with distance. 
In the massless phase, if such phase exists, one anticipates that both kind of 
charges are confined by a weakly-growing logarithmic potential, hence both 
correlations decrease with a power law. It is not obvious, however, and to the 
best of our knowledge not proved so far, that the correlation of the Polyakov 
loops and the dual correlation show similar critical behavior. In fact, as we 
shall argue below, their critical behavior is exactly opposite. Let us suppose,
there exists a massless phase and $\beta_c^{(1)}$ is the phase transition 
point from the electric confinement to the massless phase, while 
$\beta_c^{(2)}$ is the phase transition point from the massless phase to the 
deconfined phase. One then has for the Polyakov loop correlation 
the following asymptotic behavior at the critical points (for $j=1$)
\begin{equation}
P_j(R;\beta_c^{(1)};N) \ \asymp \  \frac{1}{R^{\eta^{(1)}}} \ \ , \ \ \ \
P_j(R;\beta_c^{(2)};N) \ \asymp \  \frac{1}{R^{\eta^{(2)}}} \ . 
\label{PLcorr_1}
\end{equation}
In the case of the dual correlations, the $\eta$ indices are interchanged 
\begin{equation}
\Gamma_j(R;\beta_c^{(1)};N) \ \asymp \  \frac{1}{R^{\eta^{(2)}}} \ \ , \ \ \ \ 
\Gamma_j(R;\beta_c^{(2)};N) \ \asymp \  \frac{1}{R^{\eta^{(1)}}} \ . 
\label{Corr_dual_1}
\end{equation}
If the $3D$ $Z(N>4)$ vector gauge model belongs to the universality class of 
the corresponding $2D$ $Z(N)$ spin model, we should find 
\begin{equation} 
\eta^{(1)} \ = \ 1/4 \ \ , \ \ \eta^{(2)} \ = \ 4/N^2 \ .
\label{eta_cr}
\end{equation}
For $N$ fixed and $N_t$ increasing, $\beta_c^{(1)}\to\beta_c^{(2)}$. 
In the limit $N_t\to\infty$ one ends up with a single phase transition 
from the confining to the deconfining phase. When $N_t$ is fixed and $N$ 
increases, $\beta_c^{(2)}$ diverges roughly as $N^2$, while $\beta_c^{(1)}$ 
approaches the critical point of the finite-temperature $U(1)$ LGT 
exponentially fast. 

The direct proof of all these properties would include the construction of 
RG equations describing the behavior of the system in the vicinity of phase 
transitions, something we could not accomplish so far. 
In the absence of such a proof, we give qualitative arguments why the
behavior~(\ref{PLcorr_1})-(\ref{Corr_dual_1}) is rather natural and might be 
anticipated. First of all, the interchange of the critical indices can be 
easily seen in the $2D$ $Z(N)$ spin models. Indeed, Eq.~(\ref{PLcorr_dual}) 
remains formally correct for the two-point correlation function of these models 
if one replaces the surface $S_d$ in~(\ref{source}) with a dual path 
connecting the points $0$ and $R$. The two-point correlation of dual spins can 
be written in precisely the same form of Eq.~(\ref{PLcorr_dual}) in the 
original formulation, with $Q_d(s)$ substituted by $Q(s)$. 
Let us consider now the Villain formulation. For $Q(s)$ it means 
\begin{equation} 
Q(s) \ \to \ Q^V(s) \ = \ \sum_{m=-\infty}^{\infty}
\exp \left [ -\frac{1}{2}\beta \frac{4 \pi^2}{N^2} 
\left ( s + N m \right )^2 \right ] \;,
\label{Qs_V}
\end{equation}
while for $Q_d(s)$ it amounts to replacing the Bessel function in~(\ref{Qdual})
with its asymptotics 
\begin{equation} 
Q_d(s) \ \to \ Q_d^V(s) \ = \ \sum_{m=-\infty}^{\infty}
\exp \left [ -\frac{1}{2\beta} \left ( s + N m \right )^2 \right ]  \ .
\label{Qs_V_dual}
\end{equation}
In this formulation the $2D$ $Z(N)$ model is self-dual. From this fact, from 
Eq.~(\ref{PLcorr_dual}) and its analog for the dual correlation function 
it follows that (we keep the notation $P_j$ for the correlation function of 
spins) 
\begin{equation} 
P_j(R;\beta;N) \ = \ \Gamma_j\left(R;\frac{N^2}{4\pi^2\beta};N\right) \ .
\label{corr_relation}
\end{equation}
It is now straightforward to repeat the calculations of Ref.~\cite{elitzur} 
for the dual correlation function and to get
\begin{equation} 
P_j(R;\beta;N) \ \asymp \ \exp \left [ -\frac{j^2}{2\pi \beta_{\rm eff}} 
\ \ln R \right ] \ , \ 
\Gamma_j\left(R;\frac{N^2}{4\pi^2\beta};N\right) 
\ \asymp \ \exp\left [ -\frac{2\pi j^2 \beta_{\rm eff}}{N^2} \ \ln R\right]\ . 
\label{corr_2D}
\end{equation}
$\beta_{\rm eff}$ can be expanded in powers of the self-energies of the 
topological defects of the model. The leading contribution $\beta_{\rm eff}
\approx \beta$ comes from the spin-wave configurations. At the critical points 
one has 
\begin{equation} 
\beta_{\rm eff}(\beta_c^{(1)}) = 2/\pi \ \ , \ \   
\beta_{\rm eff}(\beta_c^{(2)}) = N^2/(8\pi) \ .
\label{beta_eff}
\end{equation}
It then follows from Eqs.~(\ref{corr_2D})-(\ref{beta_eff}) that the critical 
indices are indeed interchanged. $2D$ $Z(N)$ vector models in the standard 
formulation are not self-dual, therefore Eq.~(\ref{corr_relation}) does not 
generally hold. Nevertheless, this pictures remains qualitatively correct for 
all formulations of $2D$ models which belong to the same universality class. 
Moreover, recalling that $| \beta_1 | \gg | \beta_k |$ for all 
$k\ne 1(\mod N)$, one finds the approximate equality $P_j(R;\beta;N) \approx 
\Gamma_j(R;\beta_1;N)$ for the standard formulation, where $\beta_k$ are the 
dual coupling constants~(\ref{couplings}). 

Let us turn now to the $3D$ $Z(N)$ LGT and use again the Villain formulations, 
Eqs.~(\ref{Qs_V}) and (\ref{Qs_V_dual}) (in the case of the gauge model one 
has to take the plaquette angle $s(p)$ defined in Eq.~(\ref{plaqangle})). In 
the spin-wave approximation, which should be valid in the massless phase, we 
obtain 
\begin{equation} 
P_j(R;\beta;N) \ \asymp \ \exp \left [ -\frac{N_t}{\beta} j^2 \ D(R) \right ] 
\label{corr_PL_3D}
\end{equation}
for the Polyakov loops correlation function and 
\begin{equation} 
\Gamma_j(R;\beta;N) \ \asymp \ \exp \left [ -\frac{4\pi^2\beta}{N^2} j^2 
\ G(R) \right ] 
\label{corr_dual_3D}
\end{equation} 
for the dual correlations. $D(R)$ in Eq.~(\ref{corr_PL_3D}) is the 
two-dimensional Green's function. $G(R)$ in Eq.~(\ref{corr_dual_3D}) is the 
three-dimensional Green's function. At finite temperature in the massless 
phase the dominant contribution to $G(R)$ arises from the temporal zero mode, 
other modes being massive and exponentially suppressed. We thus have 
\begin{equation*}
G(R) \ \approx \ \frac{1}{N_t} \ D(R) + {\cal{O}} (e^{-mR}) \ \ , \  \ 
D(R) \ \approx \ \ \frac{1}{2\pi} \ \ln R \ .  
\end{equation*}
Combining this with Eq.~(\ref{corr_PL_3D}) and Eq.~(\ref{corr_dual_3D}) leads 
to 
\begin{equation} 
P_j(R;\beta;N) \ \asymp \ \exp \left [ -\frac{N_t}{2\pi\beta} j^2 \ \ln R 
\right ] \ , \ \Gamma_j(R;\beta;N) \ \asymp \ \exp \left [ -\frac{2\pi\beta}
{N_t N^2}  j^2 \ \ln R \right ] \ .
\label{corr_final}
\end{equation} 
These expressions must be supplemented by computing the corrections from 
the topological defects of the model, like $Z(N)$ monopoles. We expect that
this results, similarly to $2D$ models, in replacing $\beta\to\beta_{\rm eff}$ 
in the above equations. 
It is natural to suppose that the analog of Eq.~(\ref{beta_eff}) takes the form
\begin{equation} 
\beta_{\rm eff}(\beta_c^{(1)}) = \frac{2}{\pi} \ N_t \ \ , \ \ \ \ \ 
\beta_{\rm eff}(\beta_c^{(2)}) = \frac{N^2}{8\pi} \ N_t \ .
\label{beta_eff_3D}
\end{equation}
It corresponds to approximately linear scaling of the critical points with 
$N_t$. If so, the critical indices are interchanged as it becomes evident 
after the substitution of the last equations in~(\ref{corr_final}). As a 
matter of fact, Eq.~(\ref{beta_eff_3D}) is our conjecture which remains to be 
proved. 

\section{Numerical results}

\subsection{Setup of the Monte Carlo simulation}

To study the phase transitions we used the cluster algorithm described 
in~\cite{2dzn}. We simulate the dual model defined by Eq.~(\ref{PTdual}) 
on an $N_t \times L \times L$ lattice with periodic BC. 
Simulations were carried out for $N_t = 2,\ 4$. To check directly that 
the critical indices are interchanged, as explained in the previous Section, 
we have also simulated the $Z(5)$ gauge model for $N_t=2,\ 4$ using the 
heat-bath algorithm.
As original action of the gauge model we used the conventional Wilson action.
For each Monte Carlo run the typical number of generated configurations 
was $10^6$, the first $10^5$ of them being discarded to ensure thermalization. 
Measurements were taken after 10 updatings and error bars were estimated
by the jackknife method combined with binning.

We considered the following observables:
\begin{itemize}
\item complex magnetization $M_L = |M_L| e^{i \psi}$,
\begin{equation}
\label{complex_magnetization}
M_L \ =\  \sum_{x \in \Lambda} \exp \left( \frac{2 \pi i}{N} s(x) \right) \;;
\end{equation}

\item population $S_L$
\begin{equation}
\label{population}
S_L \ =\  \frac{N}{N - 1} \left(\frac{\max_{i = 0, N - 1} n_i} {L^2 N_t} 
- \frac{1}{N} \right)\;,  
\end{equation}
where $n_i$ is number of $s(x)$ equal to $i$;

\item real part of the rotated magnetization $M_R = |M_L| \cos(N \psi)$
and normalized rotated magnetization $m_\psi = \cos(N \psi)$;

\item susceptibilities of $M_L$, $S_L$ and $M_R$:  
$\chi_L^{(M)}$, $\chi_L^{(S)}$, $\chi_L^{(M_R)}$
\begin{equation}
\label{susceptibilities}
\chi_L^{(\mathbf\cdot)} \ =\  L^2 N_t \left(\left< \mathbf\cdot^2 \right> 
- \left< \mathbf\cdot \right>^2 \right)\;;
\end{equation}

\item Binder cumulants $U_L^{(M)}$ and $B_4^{(M_R)}$,
\begin{eqnarray}
U_L^{(M)}&\ =\ &1 - \frac{\left\langle \left| M_L \right| ^ 4 
\right\rangle}{3 \left\langle \left| M_L \right| ^ 2 \right\rangle^2}\;, 
\nonumber \\
\label{binderU}
B_4^{(M_R)}&\ =\ & \frac{\left\langle \left| M_R 
- \left\langle M_R \right\rangle \right| ^ 4 \right\rangle}
{\left\langle \left| M_R - \left\langle M_R \right\rangle \right| ^ 2 
\right\rangle ^ 2 } \ . 
\label{binderBMR}
\end{eqnarray}

\end{itemize} 
The variable $s(x)$ appearing in the definitions~(\ref{complex_magnetization}) 
and~(\ref{population}) represents the dual spin in the case of simulations 
of the dual model. In the case of the gauge model, $s(x)$ is the Polyakov loop,
the sum in Eq.~(\ref{complex_magnetization}) becomes two-dimensional and 
$N_t$ must be omitted from Eqs.~(\ref{population}) 
and~(\ref{susceptibilities}). Also, when simulating the gauge model we have 
computed the average action and the specific heat in the vicinity of the 
critical points. 

\begin{figure}
\centering
\includegraphics[width=0.32\textwidth]{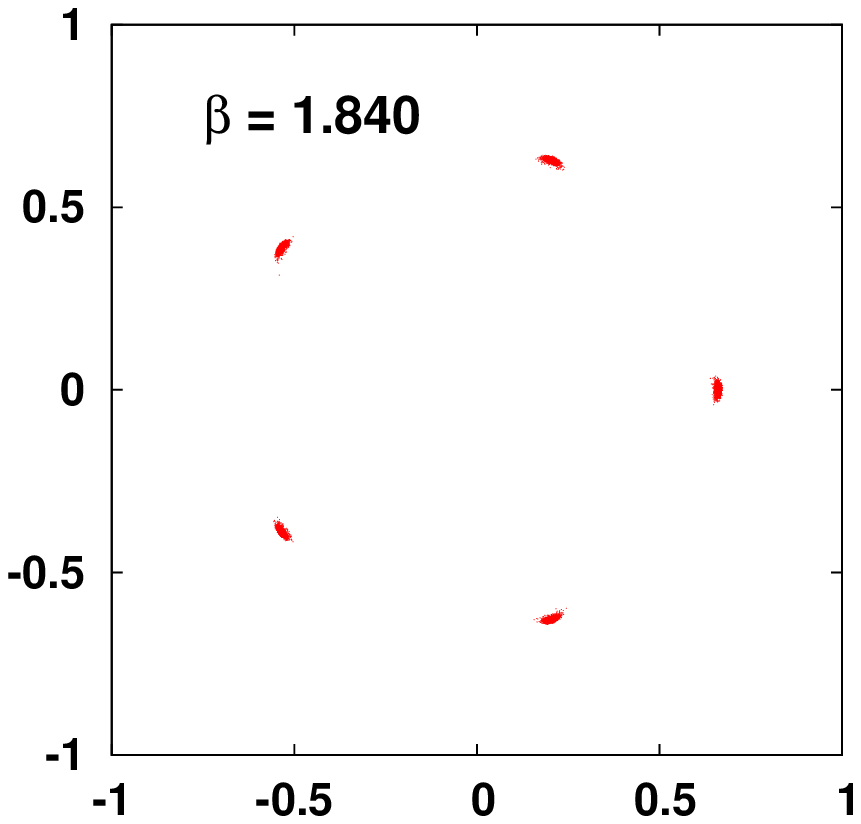}
\includegraphics[width=0.32\textwidth]{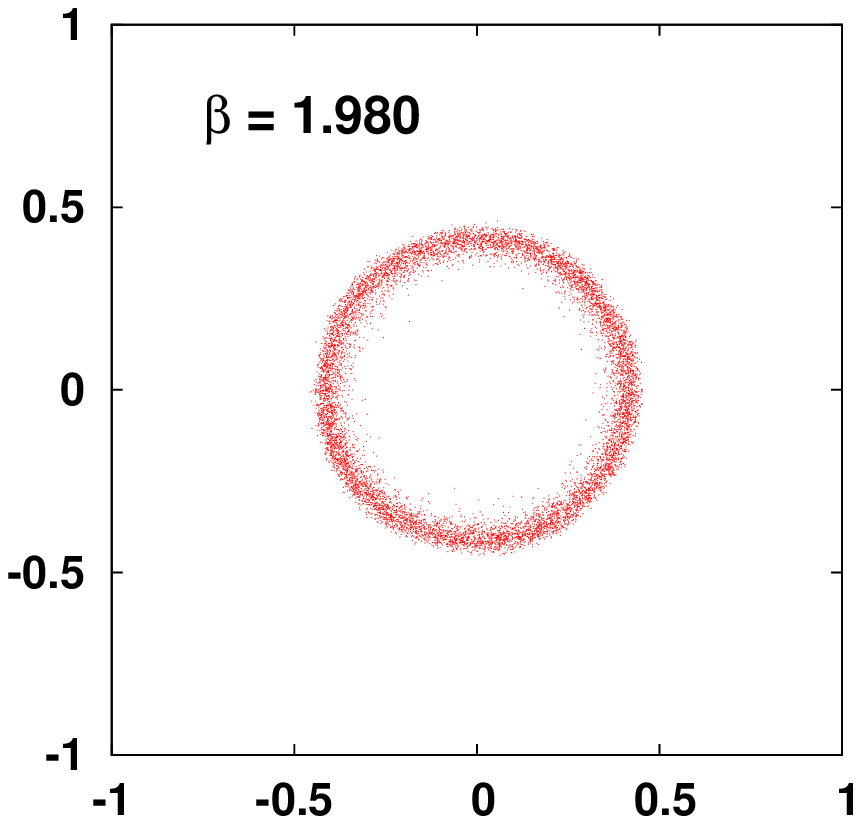}
\includegraphics[width=0.32\textwidth]{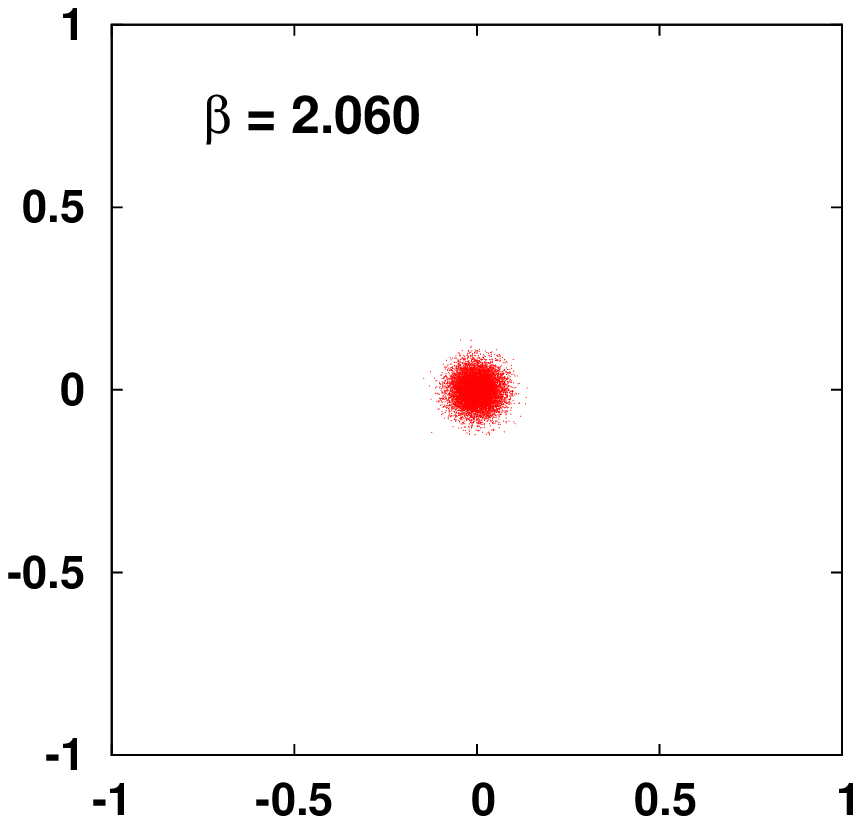}
\caption{Scatter plot of the complex magnetization $M_L$ at $\beta$=1.84, 1.98 
and 2.06 in $Z(5)$ on a $512^2\times 4$ lattice.}
\label{fig:scatter}
\end{figure}

\subsection{Determination of the critical couplings}

A clear indication of the three-phase structure emerges from the inspection
of the scatter plot of the complex magnetization $M_L$ at different values 
of $\beta$: as we move from low to high $\beta$, we observe the transition
from an ordered phase ($N$ isolated spots)  through an intermediate phase 
(ring distribution) up to the disordered phase (uniform distribution around zero). 
Fig.~\ref{fig:scatter} shows this three-phase structure for the case of 
$Z(5)$ on a $512^2\times 4$ lattice.

The first and most important numerical task is to determine the value
of the two critical couplings in the thermodynamic limit, 
$\beta_{\rm c}^{(1)}$ and $\beta_{\rm c}^{(2)}$, that separate the three 
phases. To this aim we find the value of $\beta_c$ which provides the best 
overlap of universal observables, plotted for different values of $L$ against 
$(\beta-\beta_{\rm c}^{(1)})(\ln L)^{1/\nu}$, with $\nu$ fixed at 1/2.
As these universal observables we used:
\begin{itemize}
\item Binder cumulant $B_4^{(M_R)}$ and the order parameter $m_{\psi}$ for 
the first phase transition;
\item Binder cumulant $U_L^{(M)}$ for the second phase transition.
\end{itemize}
To localize regions in which the overlap must occur, the approximate maxima of
the susceptibilities $\chi_L^{(S)}$ and $\chi_L^{(M)}$ were used.

\begin{figure}
\includegraphics[width=0.49\textwidth]{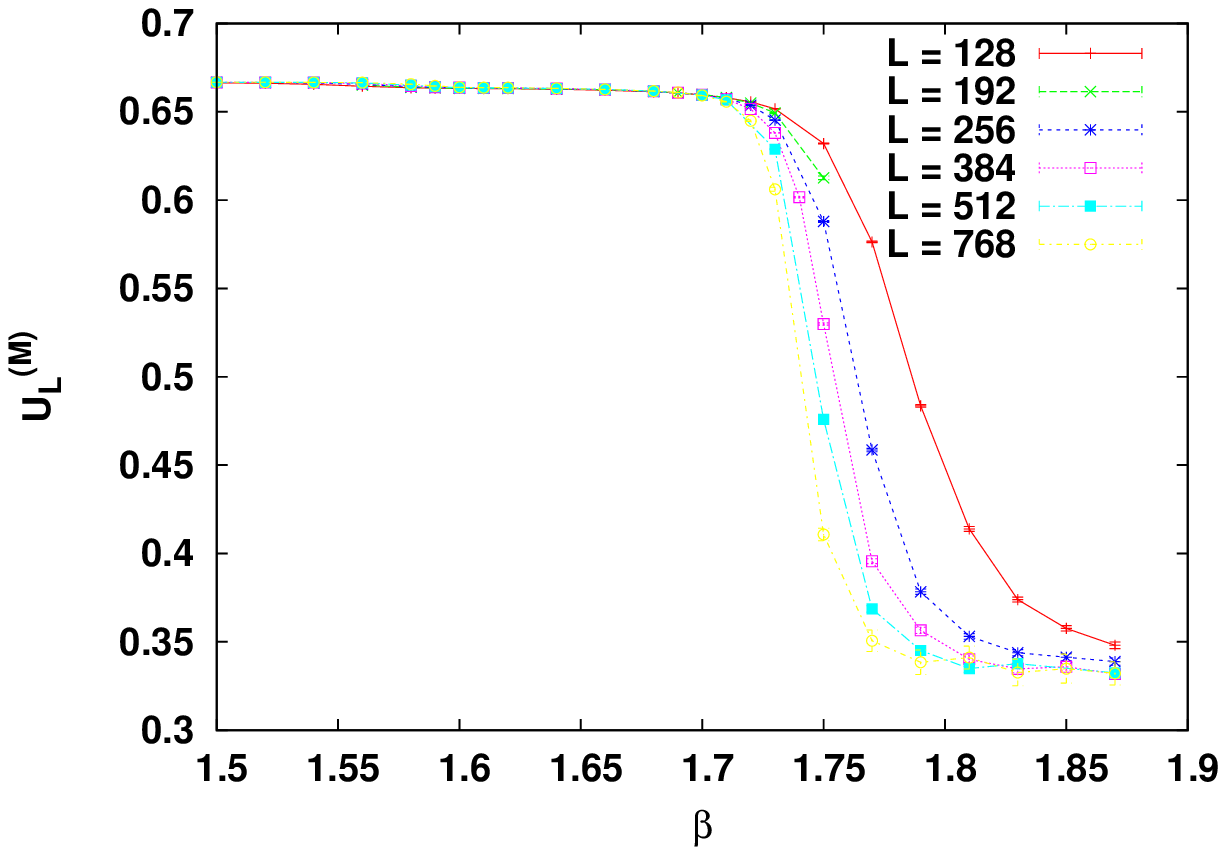}
\includegraphics[width=0.49\textwidth]{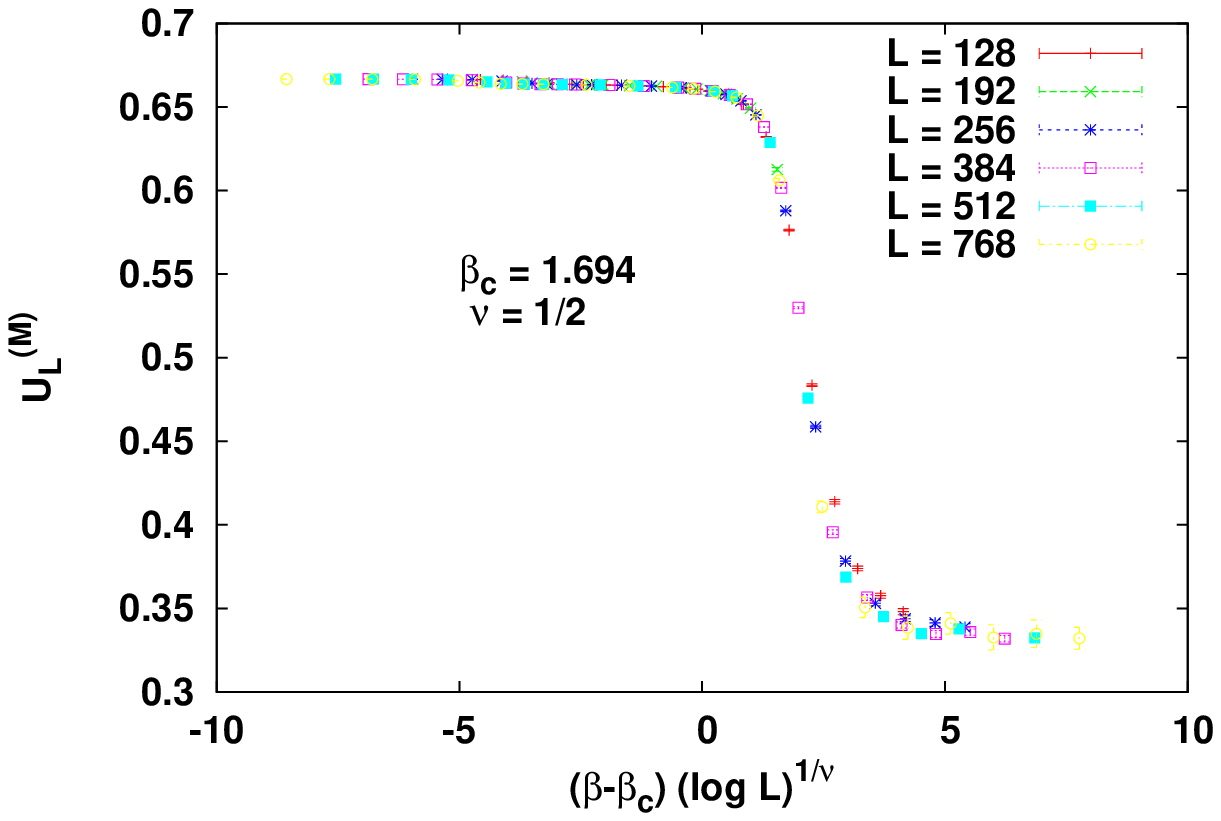} 
\caption{Binder cumulant $U_L^{(M)}$ as function of $\beta$ (left) and
 of $(\beta-\beta_c) (\ln L)^{1/\nu}$ (right) in $Z(5)$ $N_t = 2$ model.}
\label{U_plots_Z5_NT2}
\end{figure}

\begin{figure}
\includegraphics[width=0.49\textwidth]{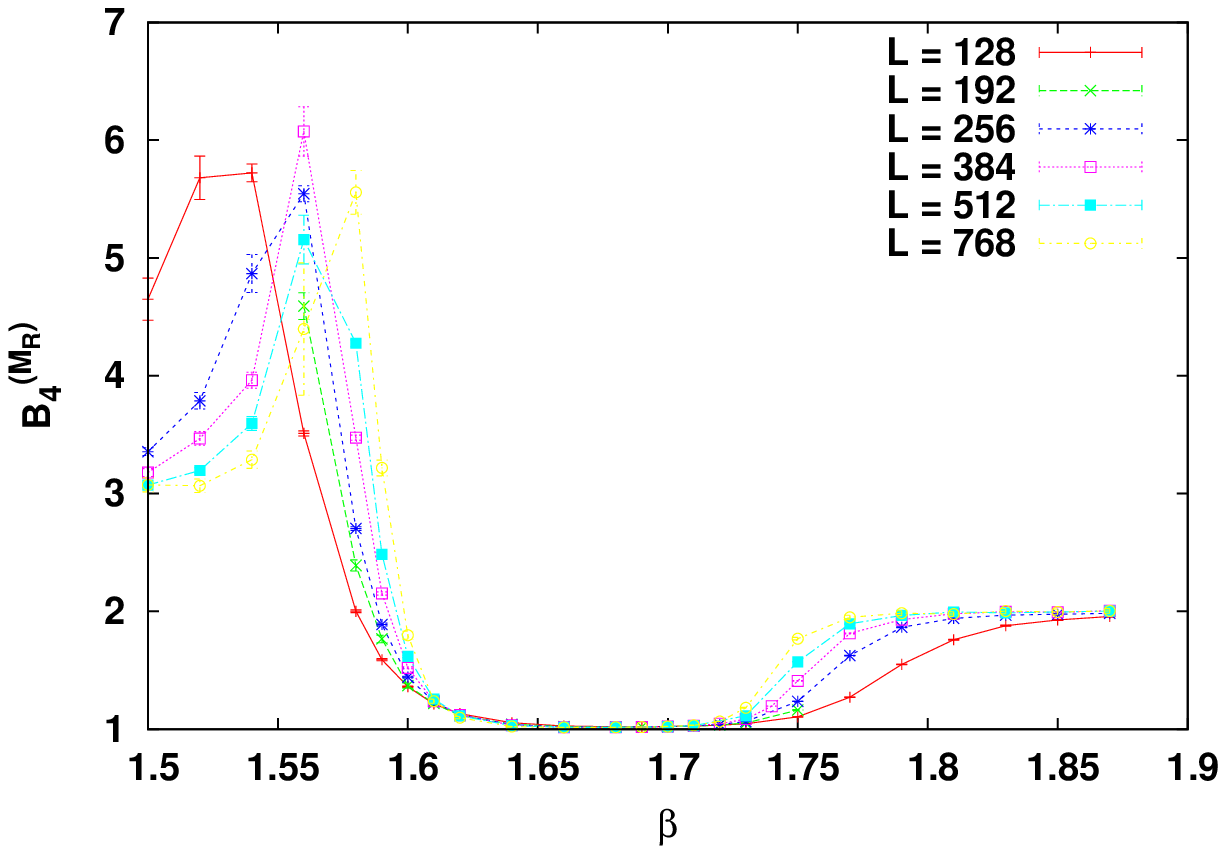}
\includegraphics[width=0.49\textwidth]{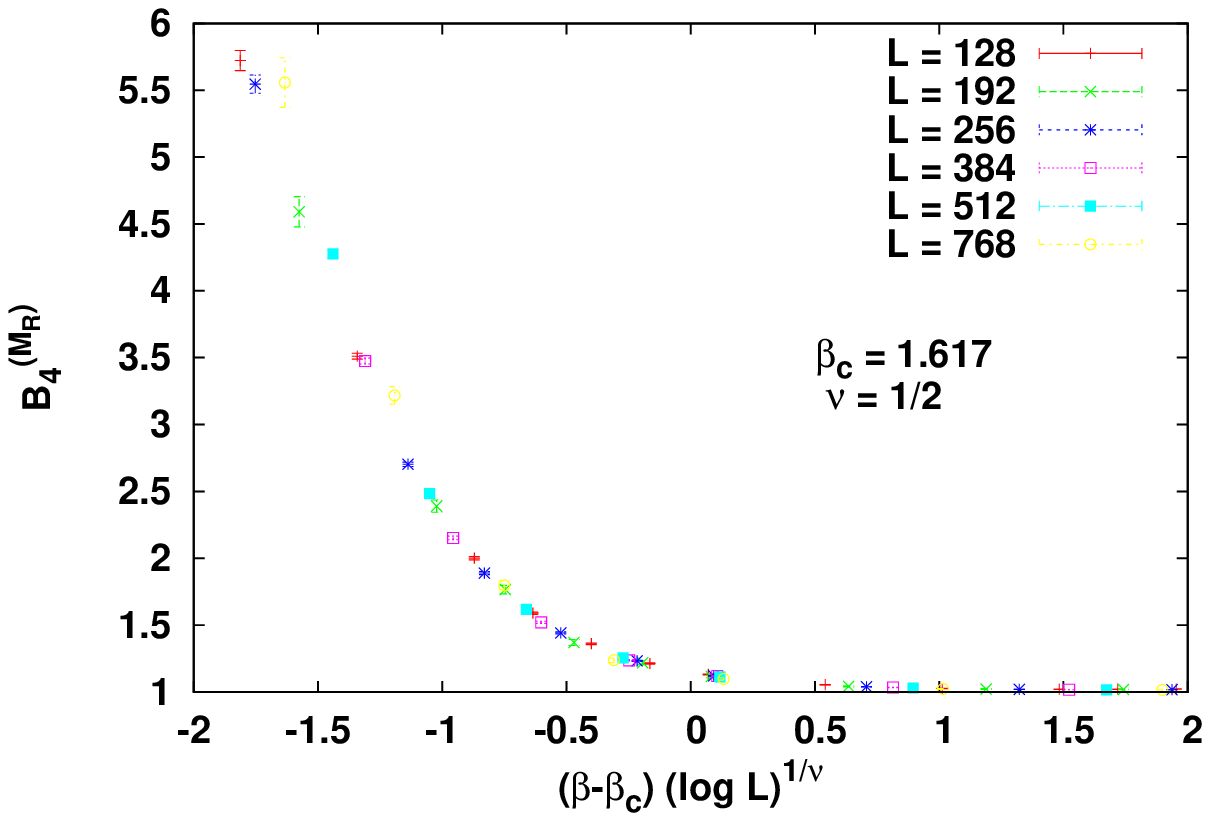}  
\caption{Binder cumulant $B_4^{(M_R)}$ as function of $\beta$ (left) and
 of $(\beta-\beta_c) (\ln L)^{1/\nu}$ (right) in $Z(5)$ $N_t = 2$ model.}
\label{B_plots_Z5_NT2}
\end{figure}

\begin{figure}
\includegraphics[width=0.49\textwidth]{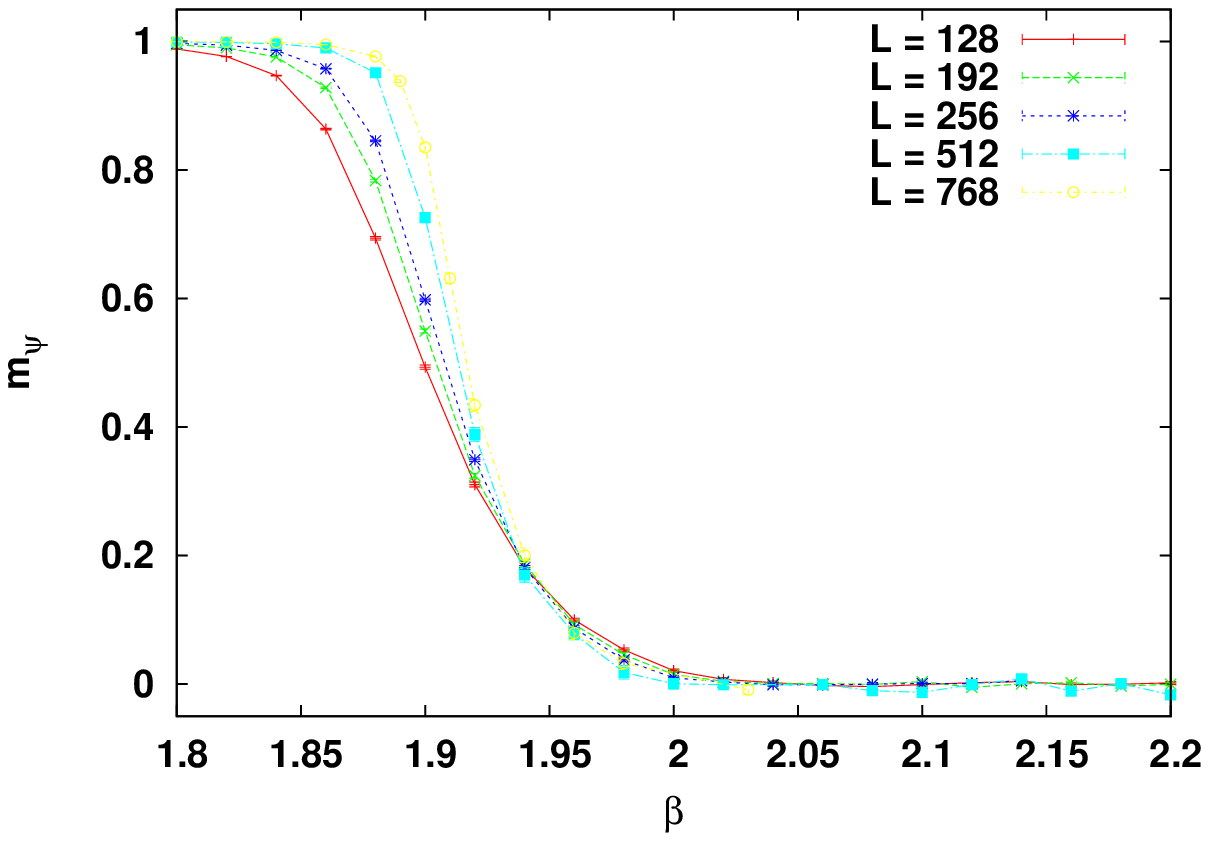}
\includegraphics[width=0.49\textwidth]{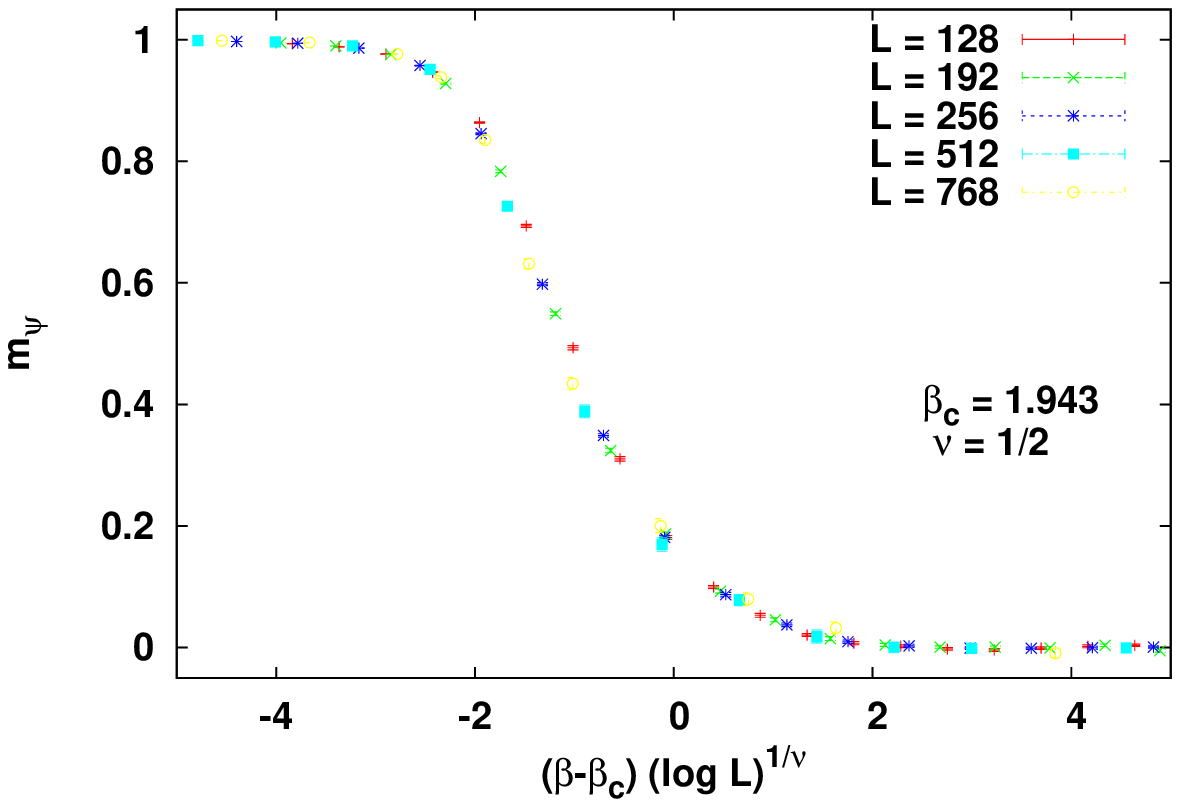} 
\caption{$m_{\psi}$ as function of $\beta$ (left) and
 of $(\beta-\beta_c) (\ln L)^{1/\nu}$ (right) in $Z(5)$ $N_t = 4$ model.}
\label{mpsi_plots_Z5_NT4}
\end{figure}

In Figures~\ref{U_plots_Z5_NT2}-\ref{mpsi_plots_Z5_NT4} we give the plots
of the universal observables, drawn against $\beta$ and against 
$(\beta-\beta_{\rm c}^{(1)})(\ln L)^{1/\nu}$.
We report in Table~\ref{tbl:crit_betas} the determinations of the
critical couplings $\beta_{\rm c}^{(1)}$ and $\beta_{\rm c}^{(2)}$
in $Z(N)$ with $N$=5 and 13 for $N_t$=2 and 4.

\begin{table}[ht]
\caption{Values of $\beta_{\rm c}^{(1)}$ and $\beta_{\rm c}^{(2)}$ obtained 
for $N_t$ = 2 and 4 in $Z(N)$ with $N = 5,\ 13$.}
\begin{center}
\begin{tabular}{|c|c|c|c|}
\hline
$N$ & $N_t$ & $\beta_{\rm c}^{(1)}$ & $\beta_{\rm c}^{(2)}$ \\
\hline
 5 & 2 & 1.617(2) & 1.694(2) \\
 5 & 4 & 1.943(2) & 1.990(2) \\ 
\hline
 13 & 2 & 1.795(4) & 9.699(6) \\
 13 & 4 & 2.74(5) & 11.966(7) \\
\hline
\end{tabular}
\end{center}
\label{tbl:crit_betas}
\end{table}

\subsection{Determination of critical indices at the two transitions}
\label{sec:betas}

Once critical couplings have been estimated, we are able to extract some
critical indices and check the hyperscaling relation. 

Since we are using the observables in the dual model the transitions change 
places: the first transition is governed by the behavior of $M_{R}$, the 
second one by the behavior of $M_{L}$. This could already be seen in the 
previous Section, since the first critical point was obtained from the
$B_4^{(M_R)}$ and the second one from the $U_L^{(M)}$ curve collapse.

We start the discussion from the second transition. According to the 
standard finite-size scaling (FSS) theory, the equilibrium magnetization 
$|M_{L}|$ at criticality should obey the relation 
$|M_{L}| \sim L^{-\beta / \nu}$, if the spatial extension $L$ of the lattice
is large enough~\footnote{The symbol $\beta$ here denotes a critical index and 
not, obviously, the coupling of the theory. In spite of this inconvenient 
notation, we are confident that no confusion will arise, since it will be 
always clear from the context which $\beta$ is to be referred to.}. Therefore,
we fit data of $|M_L|$ at $\beta^{(2)}_{\rm c}$, on all lattices with size 
$L$ not smaller than a given $L_{\rm min}$, with the scaling law
\begin{equation}
|M_{L}|=A L^{-\beta/\nu}\;.
\label{magn_fss}
\end{equation}

The FSS behavior of the susceptibility $\chi^{(M)}_L$ is given by 
$\chi^{(M)}_L\sim L^{\gamma/ \nu}$, where $\gamma/\nu=2-\eta$ and $\eta$ is 
the magnetic critical index. Therefore we fit data of $\chi^{(M)}_L$ at
$\beta^{(2)}_{\rm c}$, on all lattices with size $L$ not smaller 
than a given $L_{\rm min}$, according to the scaling law
\begin{equation}
\chi^{(M)}_{L}= A L^{\gamma/\nu}\;.
\label{chiM_fss}
\end{equation}
As the value of the critical coupling $\beta^{(2)}_{\rm c}$ we use the
central value determined in the previous Section. 

The results of the fits are summarized in Table~\ref{indices_second}. Each row 
corresponds to the fit using data from $L = L_{\min}$ up to $L = 1024$. The 
reference value for the index $\eta$ at this transition is 1/4, whereas the
the hyperscaling relation to be fulfilled is $\gamma/\nu+2\beta/\nu=d=2$.
 
\begin{table}
\caption{Critical indices $\beta/\nu$ and $\gamma/\nu$ for the second
transition in $Z(N)$ models, determined by the fits given in 
Eqs.~(\ref{magn_fss}) and~(\ref{chiM_fss}) on the complex magnetization
$M_L$ and its susceptibility $\chi_L^{(M)}$ at $\beta_{\rm c}^{(2)}$ 
for different choices of the minimum lattice size $L_{\rm min}$.
The $\chi^2$ of the two fits, given in the columns four and six, is the reduced
one.}
\begin{center}
\setlength{\tabcolsep}{2.5pt}
\begin{tabular}{|c|c|c|c|c|c|c|c|}
\hline
model & $L_{\min}$ & $\beta/\nu$  & $\chi^2_{\beta/\nu}$ 
           & $\gamma/\nu$ & $\chi^2_{\gamma/\nu}$ 
           & $d = 2\beta/\nu + \gamma/\nu$   & $\eta = 2-\gamma/\nu$ \\
\hline
                        & 128 &  0.1226(4) & 0.92  & 1.76(1)  & 1.75 & 2.00(1) & 0.24(1) \\
$Z(5)$                  & 192 &  0.1226(6) & 1.15  & 1.75(2)  & 2.14 & 2.00(2) & 0.25(2) \\
$N_t = 2$               & 256 &  0.1226(9) & 1.53  & 1.77(2)  & 1.51 & 2.02(2) & 0.23(2) \\ 
$\beta_c^{(2)} = 1.694$ & 384 &  0.121(1)  & 0.93  & 1.74(2)  & 0.78 & 1.98(3) & 0.26(2) \\ 
                        & 512 &  0.1230(2) & 0.011 & 1.74(5)  & 1.46 & 1.99(5) & 0.26(5) \\ 
\hline
                        &  32 & 0.1078(2) & 5.17 & 1.69(1)   & 51.5 & 1.91(1)  & 0.31(1) \\
                        &  64 & 0.1075(1) & 1.56 & 1.71(1)   & 19.6 & 1.93(1)  & 0.29(1) \\
$Z(5)$                  & 128 & 0.1074(1) & 1.39 & 1.734(8)  & 5.50 & 1.949(8) & 0.266(7) \\
$N_t=4$                 & 192 & 0.1075(2) & 1.67 & 1.744(7)  & 2.65 & 1.959(7) & 0.256(7) \\
$\beta_c^{(2)} = 1.990$ & 256 & 0.1077(2) & 1.27 & 1.752(6)  & 1.40 & 1.968(7) & 0.248(6) \\ 
                        & 384 & 0.1076(4) & 1.67 & 1.760(8)  & 1.10 & 1.975(9) & 0.240(8) \\ 
                        & 512 & 0.1081(5) & 1.15 & 1.77(1)   & 1.57 & 1.98(1)  & 0.23(1) \\ 
\hline
\hline
                        & 128 & 0.1225(4) & 1.03   & 1.749(8) & 0.428  & 1.994(9) & 0.251(8) \\
$Z(13)$                 & 192 & 0.1225(6) & 1.29   & 1.758(7) & 0.232  & 2.003(8) & 0.242(7) \\
$N_t=2$                 & 256 & 0.1229(8) & 1.49   & 1.749(6) & 0.126  & 1.995(8) & 0.251(6) \\ 
$\beta_c^{(2)} = 9.699$ & 384 & 0.123(1)  & 2.20   & 1.741(9) & 0.103  & 1.99(1)  & 0.259(9) \\ 
                        & 512 & 0.1203(2) & 0.0187 & 1.73(1)  & 0.0794 & 1.97(1)  & 0.27(1) \\ 
\hline
                         & 32  & 0.1266(5) & 16.62 & 1.70(1)  & 19.96 & 1.95(1)  & 0.30(1) \\ 
$Z(13)$                  & 384 & 0.1275(4) & 5.14  & 1.72(1)  & 7.70  & 1.98(1)  & 0.28(1) \\ 
$N_t=4$                  & 128 & 0.1282(2) & 0.54  & 1.747(8) & 1.58  & 2.004(8) & 0.253(8) \\
$\beta_c^{(2)} = 11.966$ & 192 & 0.1283(3) & 0.76  & 1.76(1)  & 1.54  & 2.01(1)  & 0.24(1) \\
 & 256 & 0.1282(6) & 1.45  & 1.74(2)  & 1.14  & 2.00(2)  & 0.26(2) \\ 
\hline
\end{tabular}
\end{center}
\label{indices_second}
\end{table}

We see that in most cases the values of $\eta$ and $d$ are close to those
predicted by universality. The discrepancy from the exact values 
$\eta = 0.25$ and $d = 2$ may be caused by the asymptotically vanishing parts 
of the scaling behavior of the observables $|M_L|$ and $\chi^{(M)}_L$, that 
we are not taking into account, but may be significant for smaller lattice 
sizes.

The procedure for the determination of the critical indices at the first
transition is similar to the one for the second transition, with the difference
that the fit with the scaling laws Eqs.~(\ref{magn_fss}) and~(\ref{chiM_fss}) 
is to be applied to data of the rotated magnetization, $M_R$, and of 
its susceptibility, $\chi_L^{(M_R)}$, respectively.
As the value of the critical coupling $\beta^{(1)}_{\rm c}$ we use the central 
value determined in the previous Section. 

The results of the fits are summarized in Table~\ref{indices_first}.
The reference value for the index $\eta$ at this transition is 
$4/N^2$, {\it i.e.} $\eta=0.16$ for $N=5$ and $\eta\approx 0.0237$ for $N=13$,
whereas the hyperscaling relation to be fulfilled is $\gamma/\nu+2\beta/\nu=d
=2$.

A general comment is that, in many of the cases we investigated, both $d$ and 
$\eta$ at the two critical points slightly differ from the expected values,
though these differences cancel to a large extent if we define $\eta$ as 
$2 \beta/\nu$.

\begin{table}
\caption{Critical indices $\beta/\nu$ and $\gamma/\nu$ for the first
transition in $Z(N)$ models, determined by the fits given in 
Eqs.~(\ref{magn_fss}) and~(\ref{chiM_fss}) on the rotated magnetization 
$M_R$ and its susceptibility $\chi_L^{(M_R)}$ at $\beta_{\rm c}^{(1)}$ 
for different choices of the minimum lattice size $L_{\rm min}$.}
\begin{center}
\setlength{\tabcolsep}{2.5pt}
\begin{tabular}{|c|c|c|c|c|c|c|c|}
\hline
model & $L_{\min}$ & $\beta/\nu$  & $\chi^2_{\beta/\nu}$ 
           & $\gamma/\nu$ & $\chi^2_{\gamma/\nu}$ 
           & $d = 2\beta/\nu + \gamma/\nu$   & $\eta = 2-\gamma/\nu$ \\
\hline
                        & 128 & 0.097(6) & 0.101 & 1.847(5) & 0.561  & 2.04(2) & 0.153(5) \\
$Z(5)$                  & 192 & 0.103(8) & 0.093 & 1.841(6) & 0.447  & 2.05(2) & 0.159(7) \\
$N_t=2$                 & 256 & 0.10(1)  & 0.122 & 1.850(2) & 0.038  & 2.06(3) & 0.150(2) \\ 
$\beta_c^{(1)} = 1.617$ & 384 & 0.09(2)  & 0.117 & 1.851(4) & 0.056  & 2.03(4) & 0.149(4) \\ 
                        & 512 & 0.10(3)  & 0.198 & 1.848(7) & 0.091  & 2.05(8) & 0.152(7) \\ 
\hline
                        &  32 & 0.123(6) & 1.08 & 1.8403(8) & 0.72 & 2.09(1) & 0.1596(8) \\
                        &  64 & 0.118(9) & 1.13 & 1.841(1) & 0.58  & 2.08(2) & 0.159(1) \\
$Z(5)$                  & 128 & 0.11(1)  & 1.25 & 1.841(1) & 0.70  & 2.07(3) & 0.159(1) \\
$N_t=4$                 & 192 & 0.11(2)  & 1.56 & 1.842(2) & 0.86  & 2.07(4) & 0.158(2) \\
$\beta_c^{(1)} = 1.943$ & 256 & 0.11(3)  & 2.07 & 1.842(3) & 1.14  & 2.06(6) & 0.158(3) \\ 
                        & 384 & 0.07(4)  & 1.68 & 1.836(2) & 0.22  & 1.98(8) & 0.164(2) \\ 
                        & 512 & 0.06(8)  & 3.26 & 1.837(4) & 0.42  & 2.0(2)  & 0.163(4) \\ 
\hline
\hline
                        & 128 & 0.07(5)  & 1.28 & 1.968(9) & 0.97  & 2.1(1) & 0.032(9) \\
$Z(13)$                 & 192 & 0.02(5)  & 1.16 & 1.97(1) & 1.11  & 2.0(1)  & 0.03(1) \\
$N_t=2$                 & 256 & 0.04(8)  & 1.48 & 1.97(2) & 1.43  & 2.0(2)  & 0.03(2) \\ 
$\beta_c^{(1)} = 1.795$ & 384 & 0.06(14) & 2.19 & 1.98(3) & 1.93  & 2.1(3)  & 0.02(3) \\ 
                        & 512 & 0.2(17)  & 1.08 & 1.94(5) & 1.93  & 2(3)    & 0.06(5) \\ 
\hline
$Z(13)$, $N_t=4$         & 128 & $-$0.3(1) & 0.59 & 1.977(2) & 0.21  & 1.3(3) & 0.023(2) \\
$\beta_c^{(1)} = 2.74 $ & 192 & $-$0.5(1) & 0.20 & 1.980(4) & 0.24  & 0.9(2) & 0.020(4) \\
\hline
\end{tabular}
\end{center}
\label{indices_first}
\end{table}

Also here we see a general agreement between the $\eta$ and $d$ values 
obtained and those predicted by universality. However, the expected value 
of $\beta/\nu$ in Eq.~(\ref{magn_fss}) is very small, ($2/N^2$), so 
other, asymptotically vanishing, terms can have a great impact on its
determination on finite-sized lattices. This is especially evident for $Z(13)$ 
with $N_t=4$, where $\beta/\nu$ is negative indicating that the magnetization 
$M_R$ grows with lattice size. This makes problematic the determination of 
this index from simulations on lattices used in the present work and explains 
the discrepancies with the $d=2$ value.

\begin{figure}
\includegraphics[width=0.49\textwidth]{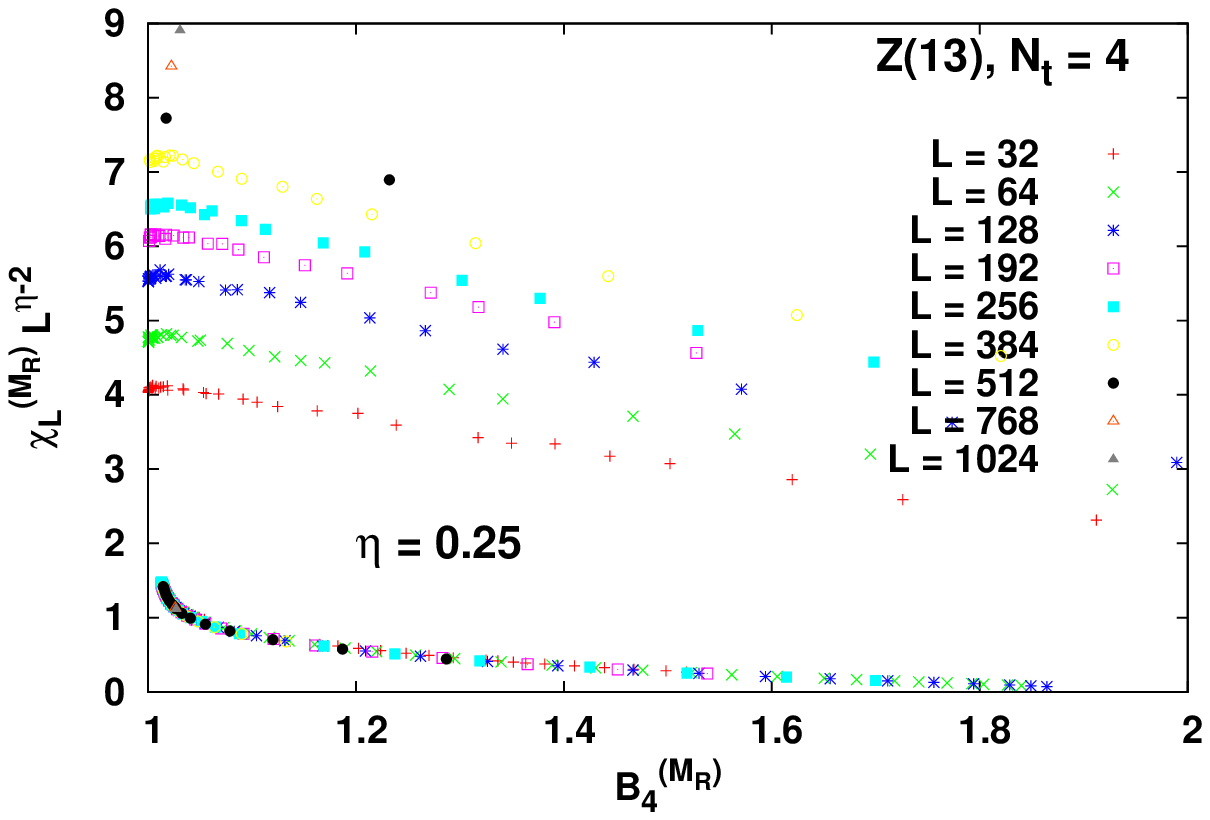}
\includegraphics[width=0.49\textwidth]{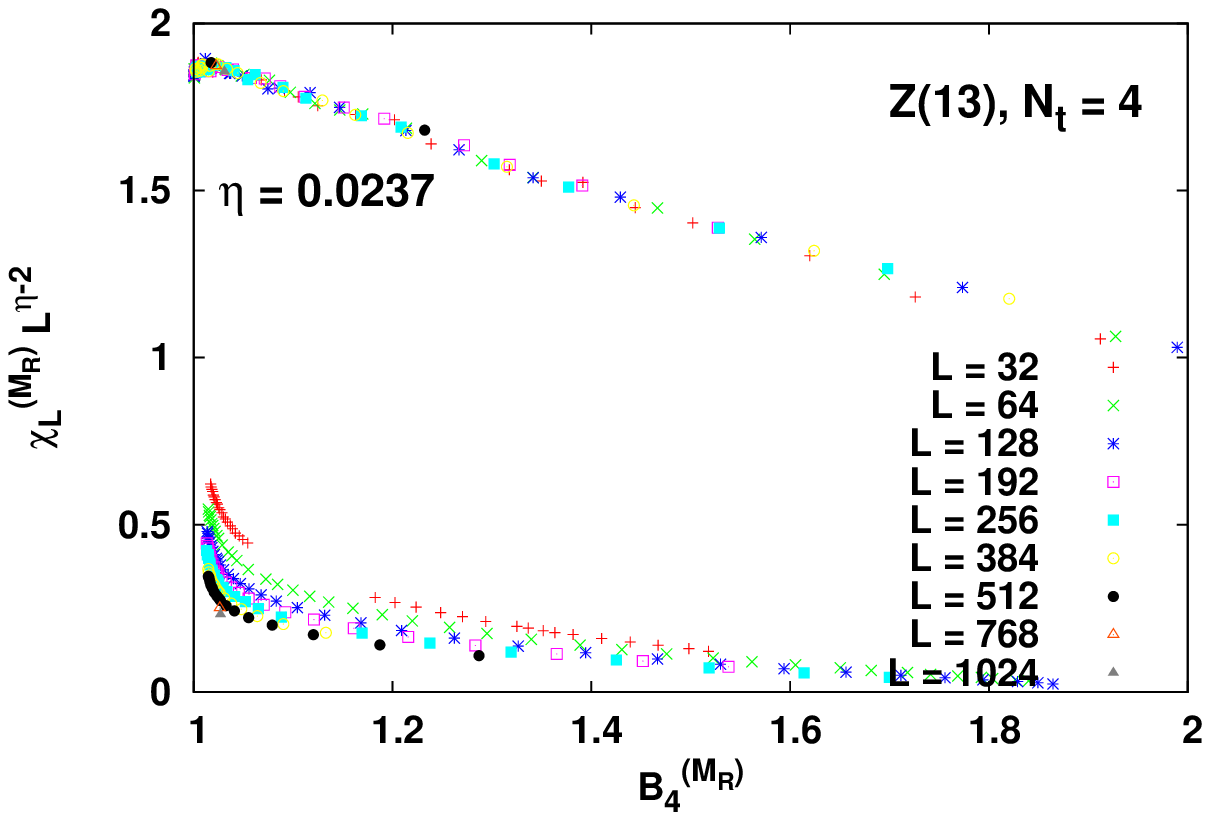}  
\caption{Correlation between $\chi_{L}^{(M_{R})}L^{\eta-2}$ and 
the Binder cumulant $B_{4}^{(M_{R})}$ in $Z(13)$ with $N_t=4$ for $\eta=0.25$ 
(left) and for $\eta=0.0237$ (right) on lattices with different size.}
\label{chimr_vs_b_Z13_NT4}
\end{figure}

\begin{figure}
\center{\includegraphics[width=0.65\textwidth]{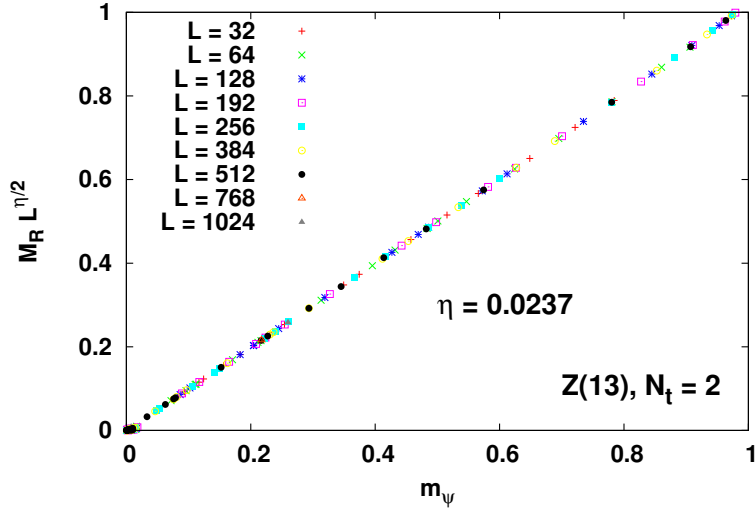}}
\caption{Correlation between $M_{R}L^{\eta/2}$ and $m_\psi$
in $Z(13)$ with $N_t=2$ for $\eta=0.0237$ on lattices with various values of 
$L$.}
\label{mrl_vs_mpsi_Z13_NT2}
\end{figure}

There is an independent method to determine the critical exponent $\eta$,
which does not rely on the prior knowledge of the critical coupling, but 
is based on the construction of a suitable universal 
quantity~\cite{Loison99,2dzn}. The idea is to plot 
$\chi_{L}^{(M_{R})}L^{\eta-2}$ versus $B_{4}^{(M_{R})}$ and to look for 
the value of $\eta$ which optimizes the overlap of curves from different 
volumes. This method is illustrated in Fig~\ref{chimr_vs_b_Z13_NT4}. 
for $Z(13)$ model with $N_t = 4$. Another option is to plot $M_{R}L^{\eta/2}$ 
versus $m_\psi$, which leads to overlapping curves for $\eta$ fixed at the 
value of the second phase transition, as illustrated in 
Fig~\ref{mrl_vs_mpsi_Z13_NT2} for $Z(13)$ model with $N_t = 2$. 

Concerning the value of the critical index $\nu$, the methods used in this work
do not allow for the direct determination of its value. When locating critical 
points we have fixed $\nu$ at 1/2. This value appears to be well in agreement 
with all numerical data. 

\subsection{Other checks of the nature of the phase transitions}

In this Section we describe briefly some results of the simulation of the
original gauge model. 
We have simulated $Z(5)$ LGT with $N_t=2,4$ and spatial extent $L\in [64-512]$.
The typical number of measurements was $10^5$. In 
Tables~\ref{indices_Z5_Nt2_gauge} and \ref{indices_Z5_Nt4_gauge} 
we present results for $N_t=2$ and $N_t=4$, correspondingly. 
In general, errors are bigger and results for critical indices are not 
so precise as in dual model simulations. Nevertheless, we can conclude 
from the inspection of Tables~\ref{indices_Z5_Nt2_gauge} 
and~\ref{indices_Z5_Nt4_gauge} that (i) the critical index $\eta$ is 
compatible with its $2D$ value and (ii) the values of the indices at two 
transitions are indeed interchanged as explained in Section 2.  

\begin{table}
\caption{Critical indices $\beta/\nu$ and $\gamma/\nu$ for the 
transitions in $Z(5)$ LGT with $N_t$ = 2.}
\begin{center}
\setlength{\tabcolsep}{2.5pt}
\begin{tabular}{|c|c|c|c|c|c|c|c|}
\hline
 & $L_{\min}$ & $\beta/\nu$  & $\chi^2_{\beta/\nu}$ 
           & $\gamma/\nu$ & $\chi^2_{\gamma/\nu}$ 
           & $d = 2\beta/\nu + \gamma/\nu$   & $\eta = 2-\gamma/\nu$ \\
\hline
                             
$\beta_{\rm c}^{(1)}= 1.617$  & 192 & 0.127(2) & 2.38 & 1.78(5) & 3.15 & 2.03(5) & 0.22(5) \\
                             
\hline
                             
$\beta_{\rm c}^{(2)}= 1.694$ & 192 & 0.1(2)  & 7.62 & 1.82(6) & 7.51 & 2.1(4) & 0.18(6) \\
\hline
\end{tabular}
\end{center}
\label{indices_Z5_Nt2_gauge}
\end{table}

\begin{table}
\caption{Critical indices $\beta/\nu$ and $\gamma/\nu$ for the 
transitions in $Z(5)$ LGT with $N_t$ = 4.}
\begin{center}
\setlength{\tabcolsep}{2.5pt}
\begin{tabular}{|c|c|c|c|c|c|c|c|}
\hline
 & $L_{\min}$ & $\beta/\nu$  & $\chi^2_{\beta/\nu}$ 
           & $\gamma/\nu$ & $\chi^2_{\gamma/\nu}$ 
           & $d = 2\beta/\nu + \gamma/\nu$   & $\eta = 2-\gamma/\nu$ \\
\hline
                             
$\beta_{\rm c}^{(1)}= 1.943$ & 128 & 0.122(4)  & 3.30 & 1.71(7) & 2.55 & 1.95(8) & 0.29(7) \\ 
                            
\hline
                             
$\beta_{\rm c}^{(2)}= 1.990$ &  64 & 0.4(2)    & 4.92 & 1.82(2)  & 1.25 & 2.6(5) & 0.18(2)  \\
\hline
\end{tabular}
\end{center}
\label{indices_Z5_Nt4_gauge}
\end{table} 

\begin{figure}
\center{\includegraphics[width=0.65\textwidth]{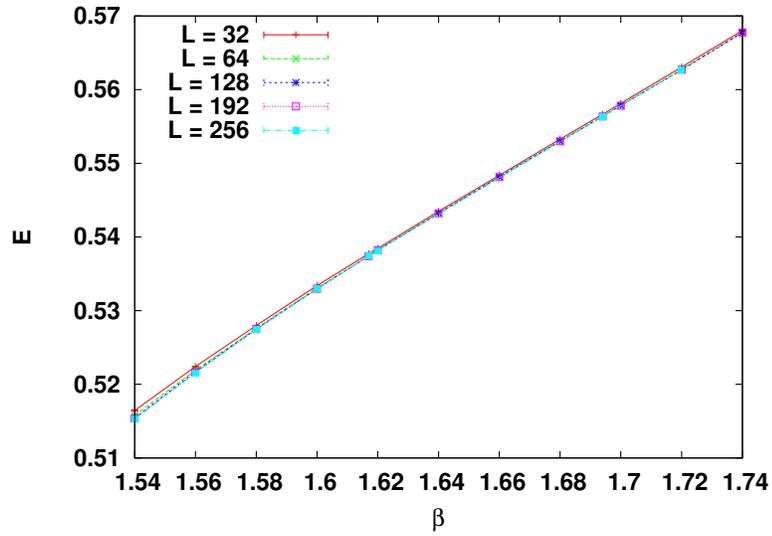}}
\caption{Average action in $Z(5)$ LGT with $N_t=2$ on lattices with various values of $L$.}
\label{fig:action_Z5_NT2}
\end{figure}

To produce further evidence in favor of the fact that the phase transitions 
investigated so far are both of infinite order, we have calculated
the average action and the specific heat around the transitions
in $Z(5)$ LGT with $N_t$=2 and $N_t=4$. In all cases the dependence 
of these quantities on $\beta$ is continuous. It follows that first and 
second order transitions are ruled out. As an example, we present  
in Fig.~\ref{fig:action_Z5_NT2} the results of simulations for 
the average action for the case of $Z(5)$ with $N_t=2$. 

\section{Conclusions and Perspectives} 

In this paper we have studied the $3D$ $Z(N)$ LGT at the finite temperature 
aiming at shedding light on the nature of phase transitions in these models 
for $N>4$.  
This study was based on the exact dual transformation of the gauge models 
to generalized $3D$ $Z(N)$ spin models. In Section~2 we presented an overview 
of the exact relation between couplings of these two models and described 
qualitatively the behavior of the Polyakov loop correlation function and 
correlation of dual spins (the disorder operator in the gauge formulation). 
Furthermore, we have advanced some arguments that the critical behavior of 
Polyakov loop correlations and dual correlations is reversed, in particular 
the values of the critical index $\eta$ at the two transitions
are interchanged for the dual correlation function with respect to the 
Polyakov loop correlation.    

The numerical part of the work has been devoted to the localization
of the critical couplings and to the computation of the critical indices. 
The main results can be shortly summarized as follows: 

\begin{itemize}

\item We have determined numerically the two critical couplings of 
$Z(N=5,13)$ LGTs and given estimates of the critical indices $\eta$
at both transitions. For the first time we have a clear indication that for 
full $3D$ $Z(N)$ vector LGT with $N\geq 5$ the scenario of three phases is 
realized: a disordered phase at small $\beta$, a massless or BKT one at 
intermediate values of $\beta$ and an ordered phase, occurring at larger and 
larger values of $\beta$ as $N$ increases.  
This matches perfectly with the $N\to\infty$ limit, {\it i.e.} the $3D$ 
$U(1)$ LGT, where the ordered phase is absent;

\item We have found that the values of the critical index $\eta$ at the two 
transitions are compatible with the theoretical expectations; 

\item The index $\nu$ also appears to be compatible with the value $1/2$, 
in agreement with universality predictions. 

\end{itemize}

On the basis of this study we are led to conclude that finite-temperature
$3D$ $Z(N)$ LGTs for $N>4$ undergo two phase transitions of the BKT type.
Moreover, these models belong to the universality class of the $2D$ $Z(N)$ 
vector models which also have two infinite order phase transitions and 
a massless phase. 

It should be emphasized that in this paper we have not studied the continuum 
limit of the theory, but rather concentrated on the very possibility of having 
two BKT-like phase transitions. At the moment we are performing simulations of 
the model on symmetric lattices with the goal to compute the zero-temperature 
string tension and extend our present computations to $N_t=8$. All these will 
help constructing the continuum limit. Also, we plan to extend our work to 
even $N$, to calculate the helicity modulus and to establish scaling formulas 
for critical points with $N$. The results of these studies will be reported 
elsewhere.

\section{Acknowledgments} 

O.B. thanks for warm hospitality the Dipartimento di Fisica dell'Universit\`a 
della Calabria and the INFN Gruppo Collegato di Cosenza during the final 
stages of this investigation. 
The work of G.C. and M.G. was supported in part by the European Union
un\-der ITN STRO\-NG\-net (grant PITN-GA-2009-238353).

\end{document}